\documentclass[12pt]{article}
\usepackage{color}

\usepackage{graphicx}
\begin{document}

\title{The role of binding entropy in the refinement of protein-ligand docking predictions: analysis based on the use of 11
scoring functions.}

\author{Anatoly~M.~Ruvinsky\thanks{Tel: 1-(785) 864-1962; Fax: 1-(785)
864-1954; E-mail address:
ruvinsky@ku.edu (A.M. Ruvinsky)} 
\\  
Center for Bioinformatics,
The University of Kansas, \\
2030 Becker Drive, Lawrence, KS 66047, USA \\
Institute of Spectroscopy, Russian Academy of Sciences,\\
Troitsk, Moscow region, 142092 Russia
}
\date{\today}
\maketitle
\newpage

\begin{abstract}

We present results of testing of the ability of eleven popular scoring functions
to predict native docked positions using a recently
developed method \cite{ruv1} for estimation the entropy
contributions of relative motions to protein-ligand
binding affinity. The method is based on the integration of 
the configurational integral over clusters obtained
from multiple docked positions. We use a test set of 100 PDB
protein-ligand complexes and ensembles of 101 docked positions
generated by Wang et al \cite{wang11} for each ligand in the test
set. To test the suggested method we compare the averaged
root-mean square deviations (RMSD) of the top-scored ligand docked
positions, accounting and not accounting for entropy
contributions, relative to the experimentally determined
positions. We demonstrate that the method increases docking
accuracy by $10-21\%$ when used in conjunction with the AutoDock
scoring function, by $2-25\%$ with G-Score, by $7-41\%$ with
D-Score, by $0-8\%$ with LigScore, by $1-6\%$ with PLP, by
$0-12\%$ with LUDI, by $2-8\%$ with F-Score, by $7-29\%$ with
ChemScore, by $0-9\%$ with X-Score, by $2-19\%$ with PMF, and by
$1-7\%$ with DrugScore. We also compare the performance of the
suggested method with the method based on ranking by cluster
occupancy only. We analyze how the choice of a RMSD-tolerance
and a low bound of dense clusters
impacts
on docking accuracy of the scoring methods. We derive optimal intervals of
the RMSD-tolerance for 11 scoring functions.\\

 \end{abstract}


\newpage
\begin{center}
\textbf{Introduction}
\end{center}

 The prediction of the
experimentally observed positions and conformations of small
organic ligands on the surface of macromolecules (e.g.
proteins, DNAs) is known as the docking problem. Methods and tools
for solving the docking problem attract a great attention in
scientific community for many years [1-13]. The accurate and fast
solution of the docking problem is of fundamental practical
importance for understanding numerous biological process in cells
and for the discovery of new drug lead compounds [3-6,11-15]. Docking
tests and detailed comparative analysis of the performance of
different docking tools [16-23] demonstrate the dependence of
docking accuracy on the conformational search methods, the quality
of the protein-ligand potentials describing binding enthalpy and
scoring methods for estimation of protein-ligand binding entropy.

Scoring functions play an important role in computational studies of protein-ligand structures and
of thermodynamics of protein-ligand binding [1,2,6-13], in virtual database screening and drug design [3-7,24-34].
We have recently suggested and validated a novel
method to estimate protein-ligand binding entropy \cite{ruv1}. We showed
that accounting for the entropy of relative and torsional motions
through a configurational integral modifies a commonly used form
of scoring functions with a term dependent on occupancy of the
clusters obtained from a number of docked positions. 
The docked
positions were generated using AutoDock \cite{auto1} docking
program and then grouped into nonoverlaping clusters in such a way
that every cluster contains ligand positions with RMSD less than a pre-set
value (a RMSD-tolerance). Ruvinsky and Kozintsev
\cite{ruv1} showed that the method essentially improves docking
accuracy in comparison with the common method based on ranking
by energy when used in conjunction with the AutoDock scoring function. 
So it is very intriguing and also important to investigate the performance of the method with 
other scoring functions.

The present article describes results of the application of
the method \cite{ruv1} in conjunction with eleven popular
scoring functions (AutoDock \cite{auto1}, G-Score \cite{gscore},
D-Score \cite{dscore}, LigScore \cite{ligscore}, PLP \cite{plp},
LUDI \cite{ludi}, F-Score \cite{fscore}, ChemScore
\cite{chemscore}, X-Score \cite{xscore}, PMF \cite{pmf}, DrugScore
\cite{drugscore}) and a test set of 100 PDB protein-ligand
complexes developed by Wang et al \cite{wang11} to predict ligand
docked positions. The test set developed by Wang et al
\cite{wang11} essentially differs from a test set of 135 PDB
complexes used by Ruvinsky and Kozintsev \cite{ruv1}. The overlap of the
test sets consists of three protein-ligand
complexes: {\it 2pk4}, {\it 1rbp} and {\it 1rnt}. Wang et al
\cite{wang11} generated ensembles of 101 docked positions for each
ligand in the test set
(http://sw16.im.med.umich.edu/software/xtool/) and scored them
by the above mentioned eleven scoring functions. Using these
ensembles
and the eleven scoring functions modified with the entropy term \cite{ruv1}, we
reordered docked ligand positions in ensembles. Then we compared the RMSD of 
top-scored ligand docked positions, accounting and not accounting for
the entropy,
 relative to the experimentally
determined positions. 

The organization of this paper is as follows. We derive
an expression for the entropy contribution of relative and torsional
motions in the Materials
and Methods section. Also in the Materials and Methods we describe
the test set of protein-ligand PDB complexes and ensembles of
docked conformations. In the Results section we compare docking
accuracies of calculations with and without entropy contributions
in terms of RMSD of the top-scored ligand docked positions
relative to the experimentally determined positions. Also in
Results we compare the suggested method with the method based on
cluster occupancy only. We summarize our conclusions in the final
section.

\begin{center}
\textbf{Materials and Methods: Theory}
\end{center}

Protein-ligand binding free energy can be written
as~\cite{mur,mici,rew} (see also [35-47])
\begin{equation}
\label{ener} \Delta G=E_{pl}-E_p-E_l-T\ln
\left(\frac{\sigma_l\sigma_p}{\sigma_{pl}}\frac{c_oN_a}{8\pi^2}\frac{Z_{pl}}{Z_pZ_l}\right),
\end{equation}
where $E_{p,l,pl}$ are the ground energies of protein (p), ligand
(l) and protein-ligand complex (pl) in solution; $N_a$ is the
Avogadro number; $c_o=1 mol/l$; $\sigma_{l,p,pl}$ are the orders
of symmetry of ligand, protein and protein-ligand complex (for a
nonsymmetrical molecule $\sigma =1$;
 if a molecule has 2-fold axis of symmetry $\sigma =2$, etc.);
$Z_{pl,p,l}$ are vibrational partition functions of proteins,
ligands and complexes. 

Considering only relative protein-ligand motions we can write the
protein-ligand binding free energy in the
form \cite{ruv1,mur,mici,rew}
\begin{equation}
\label{ener} \Delta G=E_{pl}-E_p-E_l-T\ln
\left(\frac{\sigma_l\sigma_p}{\sigma_{pl}}c_oN_a\frac{V_B}{8\pi^2}\right),
\end{equation}
where 
\begin{equation}\label{tr}
V_{B}=\int\limits_{\Gamma}\exp\left(- \frac{U_{pl}({\bf r},\theta,\varphi,\psi
)-E_{pl}}{T}\right) d{\bf r}\,\sin\theta \,d\theta\, d\varphi\,
d\psi\,
\end{equation}
is the configurational integral of the complex; 
$U_{pl}({\bf r},\theta,\varphi,\psi)$ is the energy of the
protein-ligand complex in solution; ${\bf r}$ is the vector of
relative translational motions in the complex;
$(\theta,\varphi,\psi)$ are Euler angles of relative orientational
motions; $\Gamma$ is the the region of integration in the 6-dimensional space 
of ${\bf r}$ and $(\theta,\varphi,\psi)$; $E_{pl}$ is the minimum
of $U_{pl}({\bf r},\theta,\varphi,\psi)$ in the region $\Gamma$.

Note that to predict the native binding mode corresponding to the
minimum of the Exp. (\ref{ener}), we can neglect the contribution
of $E_p+E_l$ to binding free energy. But the absolute value of the
binding constant, of course, depends on the energies of the
unbound protein and ligand molecules. Thus the binding mode is
exactly defined by $E_{pl}(\Gamma)$ and
$V_B(\Gamma)$. This property of the Exp. (\ref{tr}) essentially simplifies docking problem 
in comparison with the problem of binding affinity prediction and allows searching docked positions using probability 
distribution functions \cite{ruv2}.

Further we follow the method suggested recently \cite{ruv1}. 
In brief, it depends on the fact that
most docking algorithms generate  a number of different ligand
positions corresponding to different local minima of the
protein-ligand energy landscape. To estimate $V_B$, we first
partition all docked ligand positions (Fig.~1) from a number of
runs of an algorithm into non-overlapping clusters in such a way
that every cluster contains ligand positions with RMSD less than a
definite value ($0.5-4\AA$; see Methods section) relative to
the ligand position having minimal energy in the cluster. Now we
can consider the clusters as the possible ligand binding modes.
Further, we designate the docked ligand position having minimal
energy in the cluster as the representative position in the
cluster.
All ligand positions in the cluster numbered $i$ can be considered
as snapshots of the ligand motion near the representative docked
position $({\bf r}_i,{\bf\Omega}_i)$. The variation intervals of $({\bf r},{\bf\Omega})$ in
the cluster give the estimate of the configurational integral as
$$
V_{B}({\bf
r}_i,{\bf\Omega}_i)\approx\Gamma_i=\left[\max(\theta_i)-\min(\theta_i)\right]
[\max(\varphi_i)-\min(\varphi_i)] [\max(\psi_i)-\min(\psi_i)]
$$
\begin{equation}
\label{vb} [\max(x_i)-\min(x_i)][\max(y_i)-\min(y_i)]
[\max(z_i)-\min(z_i)],
\end{equation}
where $\max({\bf r}_i,{\bf\Omega}_i)$ and $\min({\bf
r}_i,{\bf\Omega}_i)$ are the maximum and minimum values of $({\bf
r},{\bf\Omega})$ in the cluster numbered $i$.

Omitting the contribution of $E_p+E_l$ and 
$\sigma_l\sigma_pc_oN_a/(8\sigma_{pl}\pi^2)$ 
we obtain
\begin{equation}
\label{int} \label{sc}\Delta \tilde
G_i=E_{pl}(\Gamma_i)-T\ln\Gamma_i,
\end{equation}
where $E_{pl}(\Gamma_i)$ is the energy minimum of the
protein-ligand complex in the i-mode.  To determine
the binding mode we have to determine the set $\{\Gamma\}$,
calculate the Exp. (\ref{sc}) for all $\Gamma_i$ and select a
representative position having a minimal value of $\tilde G_i$.
Exp. (\ref{int}) can also be derived by a Monte-Carlo approximation of
the configurational integral (\ref{tr})\cite{ruv1}.

The use of $\Gamma_i\approx N_iv_p$ ($v_p$
is volume per point in the configurational space, $N_i$ is the number of conformations in the cluster numbered $i$) converts Exp.
(\ref{int}) into
\begin{equation}
\label{scor} \Delta \tilde G_i=E_{pl}(\Gamma_i)-T\ln
\left(N_iv_p\right)
\end{equation}
 Thus the binding mode is
exactly defined by $E_{pl}(\Gamma_i)$ and $N_i$.
 Further,
we use the Exp. (\ref{scor}) to rank the representative positions.

To derive a scoring expression for the method of ranking by
cluster occupancy \cite{koko,ruv2,auto} (see also [50-54] and \cite{vak,cam} for
the using of the method in the studies of the
protein folding and protein-protein docking) we rewrite the Exp.
(\ref{scor}) in the following
form
\begin{equation}
\label{} P(i)=N_iv_p\exp\left(-\frac{E_{pl}(\Gamma_i)}{T}\right)
\end{equation}
$P(i)$ is proportional to the probability of finding the ligand in
the cluster i. Assuming that $E_{pl}(\Gamma)$ is a slowly
varying function over $\{\Gamma\}$, we obtain $P(i)\sim N_i$, and
the cluster occupancy becomes a single factor identifying the
native binding position. Thus the methods of ranking by cluster
occupancy or by energy $E_{pl}$ are the special cases of the more
general method based on the Exp. (\ref{scor}).

In the Results section we shall compare results of
 three scoring methods identifying the binding position   as the representative
position with the minimal value $\Delta\tilde G$ (Method 1),
 as the docked position with the minimal energy
$E_{pl}$ (Method 2), and  as the representative position in the
most occupied cluster (Method 3).

\begin{center}
\textbf{Materials and Methods: The Test Set}
\end{center}

We used the test set
 of 100 PDB protein-ligand complexes \cite{wang11}:
 1bbz, 4xia, 8xia, 2xim, 1fkf,
1fkb, 1hvr, 1tet, 2cgr, 1abf, 1apb, 7abp, 5abp, 8abp, 9abp, 1abe,
1bap, 6abp, 1e96, 1add, 2ak3, 1adb, 9aat, 1bzm, 1cbx, 2ctc, 3cpa,
1cla, 3cla, 4cla, 2csc, 5cna, 1af2, 1dr1, 1dhf, 1drf, 1ela, 7est,
3fx2, 2gbp, 1hsl, 2qwd, 2qwe, 2qwf, 2qwg, 2qwc, 2qwb, 1mnc, 1exw,
1apw, 1apt, 1bxo, 1fmo, 2pk4, 1inc, 4sga, 5sga, 5p21, 1rbp, 1rgk,
6rnt, 1rgl, 1rnt, 1zzz, 1yyy, 1b5g, 1ba8, 1bb0, 2sns, 1sre, 7tln,
4tln, 1tmn, 2tmn, 3tmn, 5tln, 1tlp, 1etr, 1ets, 1d3d, 1d3p, 1a46,
1a5g, 1bcu, 1tha, 4tim, 6tim, 7tim, 1bra, 1tnj, 1pph, 1tnk, 1tnh,
1tni, 1ppc, 1tng, 3ptb, 1tnl, 1bhf, 2xis. All these entries have
resolution better than $2.5\AA$. Wang et al \cite{wang11}
generated an ensemble of 101 docked conformations for each ligand
in the test set. One of the conformations corresponds to the
experimentally observed native conformation of the ligand. RMSD
distributions in the conformational ensembles spread from $0\AA$
to $15\AA$. For RMSD-tolerance of $2\AA$ ensembles consist of
$30-70$ distinctive conformational clusters.

To
analyze an ability of the scoring functions to predict the
experimentally observed conformations
eleven scoring functions were applied to score the conformational
ensembles \cite{wang11}. We used the scored
ensembles of each ligand and applied the Exp. (\ref{scor}) to test
the ability of the suggested method to predict the native
conformation. We varied the RMSD-tolerance from $0.5$ to $4\AA$.

\begin{center}
\textbf{Results}
\end{center}

For correct use of the Exp.~(\ref{scor}) it is necessary to keep
in mind that it is based on the estimate of the configurational integral. 
Thus only clusters with high occupancy should be scored
with Exp.~(\ref{scor}). To differentiate between dense and sparse
clusters, we introduce  a low bound $N_{lb}$ of dense clusters.
Only clusters with $N_i\ge N_{lb}$ are scored with
Exp.~(\ref{scor}). If all clusters have occupancy lower than
$N_{lb}$, we select the most occupied cluster as the cluster of
the binding position, but if several clusters have the same
occupancy lower than $N_{lb}$, we compare them using
Exp.~(\ref{scor}). Further we consider the percentage of the
top-ranked solutions within a RMSD of $2\AA$ of the experimental
result and designate this value as the success rate (SR). Success rates
of three scoring methods used in conjunction with 11 scoring
functions are given in Fig. 2-12 for the low bound of dense clusters equal to $3, 4, 5$ and $10$.

\begin{center}
\textbf{Force field based scoring functions}\\
\textbf{AutoDock}
\end{center}

As the RMSD-tolerance increases from $1\AA$ to $4\AA$, Method 1
based on Exp. (\ref{scor}) applied with the AutoDock scoring
function improves the success rate by $3-13\%$ relative to the
results of ranking by energy (Method 2) not accounting for the
entropy effect (Fig. 2-a).
SR of the Method 1 reaches
maximum of $75\%$ for the RMSD-tolerance of $2.5\AA$ and the low
bound of $4$. SR of ranking by cluster occupancy (Method 3)
reaches maximum of $73\%$ for the RMSD-tolerance of $2.5\AA$. For
the RMSD-tolerance of $0.5, 1, 3.5$ and $4\AA$ ranking by cluster
occupancy is less successful than ranking by energy.

\begin{center}
\textbf{G-Score}
\end{center}

Fig. 2-b
 shows that Method 1 applied together with the
G-Score scoring function improves the success rate by $2-25\%$
relative to the results of "bare" G-Score (Method 2). SR of the
Method 1 reaches maximum of $67\%$ for the RMSD-tolerance of
$2.5\AA$ and the low bound of $5$ and $10$. SR of ranking by
cluster occupancy reaches maximum of $70\%$ for the RMSD-tolerance
of $2.5\AA$ also. For all values of the RMSD-tolerance energy
ranking is a worse predictor than ranking based on Exp.
(\ref{scor}) and by cluster occupancy.

\begin{center}
\textbf{D-Score}
\end{center}

Applying Exp. (\ref{scor}) in conjunction with the D-Score scoring
function improves essentially the success rate by $7-41\%$
relative to the results of the Method 2 (Fig. 2-c)
. SR
of the Method 1 reaches maximum of $67\%$ for the RMSD-tolerance
of $2.5\AA$ and the low bound of $5$ and $10$. SR of the Method 3
has a maximum of $69\%$ for the RMSD-tolerance of $2.5\AA$. The
worst results of scoring ($SR=26\%$) for all values of the
RMSD-tolerance are obtained for Method 2.

\begin{center}
\textbf{Empirical scoring functions}\\
\textbf{LigScore}
\end{center}

We found that as the RMSD-tolerance increases from $2\AA$ to
$4\AA$, Method 1 applied with the LigScore scoring function
improves the success rate by $0-8\%$ relative to the results of
Method 2 (Fig. 2-d).
For the RMSD-tolerance $<2\AA$, the difference between Method 2 using
"bare"
LigScore and Methods 1 and 2 is $1-8\%$ and $6-13\%$. SR of
Method 1 using Exp. (\ref{scor}) reaches maximum of $82\%$ for the
RMSD-tolerance of $2.5\AA$ and the low bound of $3$ and $4$. SR of
Method 3 using cluster occupancy reaches maximum of $74\%$ for
the RMSD-tolerance of $2\AA$ and $2.5\AA$. For the RMSD-tolerance
$>2.5\AA$, Method 2 with energy ranking outperforms
Method 3 by $1-8\%$, but worse than Method 1 by
$0-8\%$.

\begin{center}
\textbf{PLP}
\end{center}

Docking accuracy of Method 1 applied with PLP scoring function
(Fig. 2-e)
is comparable with the accuracy of Method 2
using energy ranking for the case of the RMSD-tolerance lower than
$2\AA$,  but becomes better by $1-6\%$ for the case of the
RMSD-tolerance higher than $2\AA$. SR of the Method 1 reaches
maximum of $82\%$ for the RMSD-tolerance of $3.0\AA$ and the low
bound of $4$. SR of Method 3 using ranking by cluster
occupancy reaches maximum of $79\%$ for the RMSD-tolerance of
$3.0\AA$. For two values of the RMSD-tolerance of $3\AA$ and
$3.5\AA$ ranking by cluster occupancy slightly outperforms energy
ranking by $3\%$ and $1\%$, but loses $0-3\%$ to the results of
ranking by Method 1. For other values of the RMSD-tolerance
Methods 1, 2 show better docking results than Method 3.

\begin{center}
\textbf{LUDI}
\end{center}

Fig. 2-f
shows that as the RMSD-tolerance increases from
$0.5\AA$ to $4\AA$, Method 1 applied with the LUDI scoring
function improves the success rate by $0-12\%$ relative to the
results of Method 2. SR of Method 1 using Exp.
(\ref{scor}) reaches maximum of $79\%$ for the RMSD-tolerance of
$3.0\AA$ and the low bound of $10$. SR of Method 3 reaches
maximum of $77\%$ for the RMSD-tolerance of $3.0\AA$. Method 2
outperforms by $2-7\%$ Method 3 only in two cases of the
RMSD-tolerance of $0.5,1.0\AA$.

\begin{center}
\textbf{F-Score}
\end{center}

The results of ranking docked positions on the basis of Methods 1,
2 and 3 in conjunction with F-Score scoring function are shown in
Fig. 2-g.
We can see that as the RMSD-tolerance
increases from $1.5\AA$ to $4\AA$, Method 1 improves the
success rate by $2-8\%$ relative to the results of Method 2.
SR of Method 1 reaches maximum of $82\%$ for the
RMSD-tolerance of $3.0\AA$ and the low bound of $3,5,10$. SR of
Method 3 using ranking by cluster occupancy reaches maximum of
$77\%$ for the RMSD-tolerance of $3.0\AA$ and $3.5\AA$. For the
RMSD-tolerance lower than $2\AA$ Method 2 ($SR=74\%$)
outperforms  Method 3 by $3-9\%$.

\begin{center}
\textbf{ChemScore}
\end{center}

Fig. 2-h
shows that there is a significant
improvement of docking accuracy for Methods 1 and 3 applied
with ChemScore scoring function, in comparison with the results of
the Method 2, choosing the energy top-ranked position to identify
the binding position. Methods 1 and 3 outperform the Method 2 by
$6-29\%$ and $8-32\%$ accordingly. SR of Method 1 using Exp.
(\ref{scor}) reaches maximum of $64\%$ for the RMSD-tolerance of
$1.5, 2.0\AA$ and the low bound of $4,5$. SR of Method 3 reaches
maximum of $67\%$ for the RMSD-tolerance of $2.0\AA$.

\begin{center}
\textbf{X-Score}
\end{center}

Fig. 2-i
shows that Methods 1 and 3 applied with the
X-Score scoring function improve the success rate by $0-9\%$ for
the RMSD-tolerance from $1.0\AA$ to $4\AA$, and $0-8\%$ for the
RMSD-tolerance from $1.5\AA$ to $4\AA$ relative to the results of
Method 2 not accounting for the entropy effect. SR of 
Method 1 reaches maximum of $75\%$ for the RMSD-tolerance of
$2.5\AA$ and the low bound of $4,5,10$. SR of Method 3 using
ranking by cluster occupancy reaches maximum of $74\%$ for the
RMSD-tolerance of $2.5\AA$. For the RMSD-tolerance of $0.5, 1$ and
$4\AA$ energy ranking ($SR=66\%$) slightly outperforms by $1-3\%$
results of Method 3.

\begin{center}
\textbf{Knowledge-based scoring functions}\\
\textbf{PMF}
\end{center}

Fig. 2-j
illustrates results of applying ranking Methods
1,2 and 3 in conjunction with the PMF scoring function. We observe
that Methods 1 and 3 outperform by $2-19\%$ and $2-20\%$ the results
of the common Method 2 using energy ranking ($SR=52\%$). SR of 
Method 1 using Exp. (\ref{scor}) reaches maximum of $71\%$ for the
RMSD-tolerance of $2.5\AA$ and the low bound of $5$. SR of 
Method 3 using ranking by cluster occupancy reaches maximum of
$72\%$ for the RMSD-tolerance of $2.5\AA$.

\begin{center}
\textbf{DrugScore}
\end{center}

As the RMSD-tolerance increases from $1.0\AA$ to $4\AA$, 
Method 1 applied with the DrugScore scoring function improves the
success rate by $1-7\%$ (Fig. 2-k) relative to the results of Method 2
($SR=72\%$). SR of using Method 1 reaches maximum of $79\%$
for the RMSD-tolerance of $3.0\AA$ and the low bound of $4$, and
the RMSD-tolerance of $4.0\AA$ and the low bound of $10$. SR of
Method 3 reach maximum of $75\%$ for the RMSD-tolerance of
$2\AA$ and $3 \AA$. For the RMSD-tolerance lower than $1.5\AA$ and
equal to $4\AA$ energy ranking (Method 2) outperforms by $2-7\%$
results of Method 3 using ranking by cluster occupancy.

\begin{center}
\textbf{Discussions}
\end{center}

The success rate of ranking using Exp. (\ref{scor}) (Method 1) shows a
bell-shape curve behavior (Fig. 2) for all scoring functions except LigScore
(Fig. 2-d).
It means that
10 of 11 scoring functions closely describe protein-ligand energy landscapes in
the test
set.
Tails of the bell-shape functions approach a value of the success rate
of energy ranking neglecting the entropy effect. This is the
result of reduction of the suggested method to the method of
ranking by energy for a very small or large RMSD-tolerance of
cluster size. Indeed, for a very low RMSD-tolerance all clusters contain a
single docked position  and ranking using the Method 1 and ranking by energy
(Method 2) are became identical. In the limit of a very large
RMSD-tolerance, only one cluster exists and thus ranking by Method 1 and
Method 2 give the same results for the success rate. The bell-shape curve
behavior for PMF  is not so evident (Fig. 2-j) as for
AutoDock (Fig. 2-a), D-Score (Fig.
2-c), LUDI (Fig. 2-f) or ChemScore (Fig. 2-h). However, it can be detected 
averaging SR over different values of the low bound of the
cluster size for every value of the RMSD-tolerance or by following
rhombus and stars on Fig. 2-j. The SR behavior for Method 1 applied in
conjunction with LigScore as a function of the RMSD-tolerance and the low bound
of the cluster size  has the same character for a very low and large values
of the RMSD-tolerance, but differs in the most interesting range of
intermediate
values of the RMSD-tolerance. It has two clear extrema - a minimum for the
RMSD-tolerance of $1\AA$ and a maximum for the RMSD-tolerance of $2.5\AA$.

It is interesting to note that all scoring functions demonstrate even
behavior of a maximum as a function of the RMSD-tolerance. Thus applying 
Method 1 we can vary the RMSD-tolerance in pre-set intervals keeping a
level of docking accuracy averaged over the low bound of the cluster size. So
AutoDock allows one to vary the RMSD-tolerance from $1.5\AA$ to $3\AA$ (Fig.
2-a),
G-Score - from $1.5\AA$ to $2.5\AA$ (Fig. 2-b), D-Score - from $1.5\AA$ to
$2.5\AA$ (Fig. 2-c), LigScore - from $2.5\AA$ to $4\AA$ (Fig. 2-d), PLP -
from $2.5\AA$ to $4\AA$ (Fig. 2-e), LUDI - from $2\AA$ to $3\AA$ (Fig. 2-f),
F-Score - from $2.5\AA$ to
$4\AA$ (Fig. 2-g),
ChemScore - from $1.5\AA$ to $2.5\AA$ (Fig. 2-h),  X-Score - from $2\AA$ to
$3\AA$ (Fig. 2-i), PMF - from
$1.5\AA$ to $3.5\AA$ (Fig. 2-j), DrugScore - from $2.5\AA$ to $4\AA$ (Fig. 2-k).
Considering optimal values of the RMSD-tolerance and the low bound of the
cluster size, we found that Method 1 outperforms Method 2 in docking
accuracy by $10-21\%$ when used in conjunction with the AutoDock scoring
function, by $2-25\%$ with G-Score, by $7-41\%$ with
D-Score, by $0-8\%$ with LigScore, by $1-6\%$ with PLP, by
$0-12\%$ with LUDI, by $2-8\%$ with F-Score, by $7-29\%$ with
ChemScore, by $0-9\%$ with X-Score, by $2-19\%$ with PMF, and by
$1-7\%$ with DrugScore.
These results are the unambiguous evidence of improving docking
accuracy by accounting for the entropy of relative and torsional
motions.

The success rate of scoring over cluster occupancy (Method 3) also shows
the bell-shape curve behavior for AutoDock (Fig. 2-a), G-Score (Fig. 2-b),
D-Score (Fig. 2-c), ChemScore (Fig. 2-h), X-Score (Fig. 2-i) and PMF (Fig.
2-j). For PLP (Fig. 2-e), LigScore (Fig. 2-d), LUDI (Fig. 2-f), F-Score (Fig.
2-g), DrugScore (Fig. 2-k) curves of SR as a function of the RMSD-tolerance
differ from the bulb function behavior and show several extrema. 
Considering optimal values of the RMSD-tolerance, we found that 
Method 3 outperforms Method 2 in docking
accuracy by $11\%$ when used in conjunction with the AutoDock scoring
function, by $28\%$ with G-Score, by $43\%$ with
D-Score, by $3\%$ with PLP, by
$10\%$ with LUDI, by $3\%$ with F-Score, by $32\%$ with
ChemScore, by $8\%$ with X-Score, by $20\%$ with PMF, and by
$3\%$ with DrugScore. Best results of Method 3 applied with LigScore
coincide with the results of ranking by LigScore energy (Method 2). 

It is
interesting to note that SRs of ranking using Method 3 and Method 1
show good correlation for AutoDock, D-Score, G-Score, LUDI, ChemScore and
X-Score. It means that using these scoring functions 
and applying Methods 1 or 3 we choose the same top-scored representative
position satisfying simultaneously to the following inequalities
\begin{equation}
\label{ke}
N_1 > N_i \mbox{  and  }
N_1\exp\left(-\frac{E_{pl}(\Gamma_1)}{T}\right)>N_i\exp\left(-\frac{E_{pl}
(\Gamma_i)}{T} \right), 
\end{equation}
where 1 is the cluster number of the top-scored representative
position. Other representative positions are numbered $i\ne 1$. Using Ineq.
(\ref{ke}) we obtain
\begin{equation}
\label{q}
E_{pl}(\Gamma_1)-E_{pl}(\Gamma_i)<T\ln\frac{N_1}{N_i}
\end{equation}
On condition $E_{pl}(\Gamma_1)>E_{pl}(\Gamma_i)$, top-scored
representative positions reside not in deepest energy wells. Using Ineq.
 (\ref{q}) we can estimate the maximal difference between depths of energy wells
in protein-ligand energy landscape as $T\ln\left( N_1/\min(N_i)\right)= 
2.8\mbox{kcal/mol}$ for $T=300\mbox{K}$, $N_1=99$ and $\min (N_i)=1$. 
If $E_{pl}(\Gamma_1)<E_{pl}(\Gamma_i)$, then top-scored representative
positions reside in the deepest and mostly occupied energy wells.

\begin{center}
\textbf{Conclusions}
\end{center}

We presented results of testing 11 popular scoring functions on
100 protein-ligand complexes using the recently suggested method
\cite{ruv1}, accounting for binding entropy of relative motions in a protein-ligand complex, and two other
commonly used methods of ranking by energy or cluster occupancy.
We rigorously showed that both methods of ranking by cluster
occupancy and by energy are the special cases of the more general
method accounting for binding entropy. We applied the three
ranking methods to the conformational ensembles generated by Wang
et al \cite{wang11} and compared efficiencies of the methods in
terms of the percentage of the top-ranked solutions within a RMSD
of $2\AA$ of the experimental result (the success rate).

We demonstrated that the method based on Exp. (\ref{scor})
predicts native position significantly better than the top
energy ranking, when used in conjunction with
D-Score, ChemScore, AutoDock, G-Score, or PMF, and moderately or
slightly better when used with LigScore, PLP, LUDI, F-Score, X-Score, or
DrugScore. The presented results prove that the method can be
applied together with all types of current force fields, empirical
scoring functions or knowledge-based potentials.

For the most of tested scoring functions
we observed strong correlations between docking accuracies of two
methods of ranking using Exp. (\ref{scor}) and ranking by cluster
occupancy. These correlations suggest that for these potentials
the near-native conformations, in comparison with far-native ones, have
the greatest number of neighboring conformations within a
RMSD-tolerance. Similar trends were observed recently in
studies of protein-ligand docking \cite{koko,ruv2,auto}, predictions of
protein-protein complexes \cite{vak,cam} and studies of protein folding
landscape \cite{baker, baker2,lele,skol}. Also this concept was used by Xiang
et al \cite{honig} for loop prediction. The
authors suggested to rank conformations by a standard energy term together
with a
RMSD-dependent term that favors conformations that have many
neighbors in configurational space. 
We believe that the method to
treat the entropy effect using Exp. (\ref{scor}) should give
statistical-thermodynamic explanations of these results and prove
useful for future studies of protein folding and protein-protein
docking. 

\begin{center}
\textbf{Acknowledgements}
\end{center}

I thank to I.A. Vakser for careful review of the
manuscript.


\newpage
\begin{center}
{\bf\large Legend to figures.}
\end{center}

1. The clustering scheme. Small circles are
local minima of the protein-ligand energy landscape found in
docking.$\phantom{jhjhj}$\\

2. Percentage of the top-ranked representative solutions
within a RMSD of $2\AA$ from the experimentally determined
position, scored using Exp. (\ref{scor}) and a) AutoDock, b) G-Score,
c) D-Score, d) LigScore, e) PLP, f) LUDI, g) F-Score, h) ChemScore, i) X-Score,
j) PMF, k) DrugScore, for the low bound of dense clusters equal to 3
(circles), 4 (triangles), 5 (rhombus) and 10 (stars) as a function
of the RMSD-tolerance. Solid line corresponds to the success rate
of the scoring function neglecting the entropy effect.
Rectangles connected with by dash line correspond to the success
rate of ranking by cluster occupancy.

\newpage

\begin{center}
\includegraphics[scale=1.0]{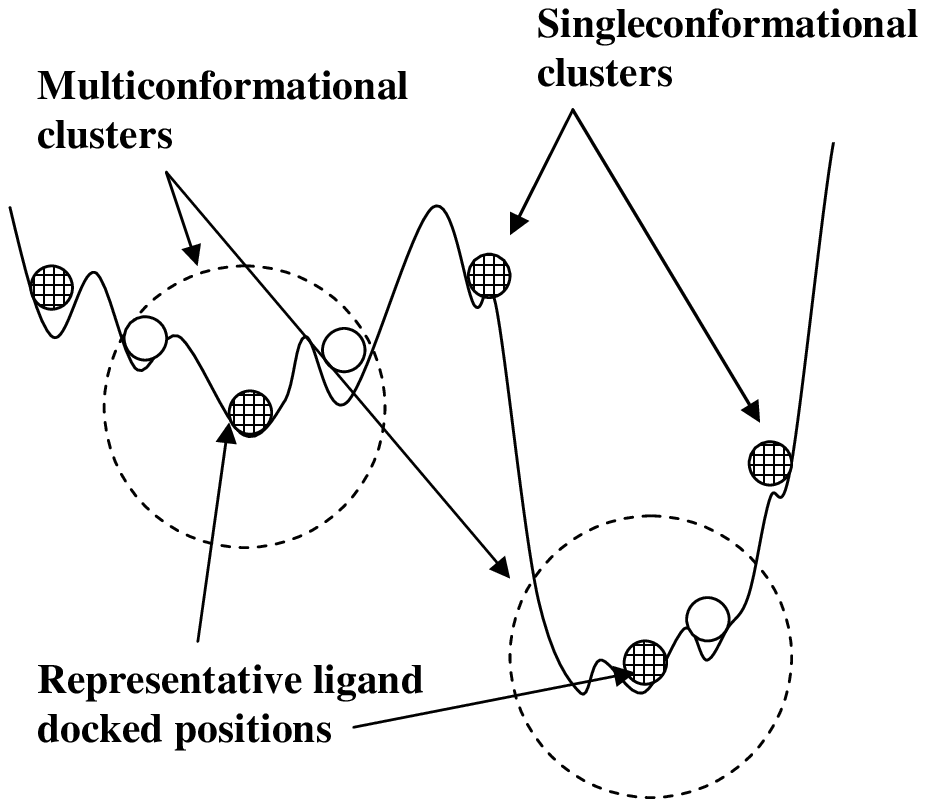}
\end{center}
\begin{center}
$\phantom{jhjhj}$\\
\textbf{\large Fig. 1.}
\end{center}


\newpage

\includegraphics[scale=1.7]{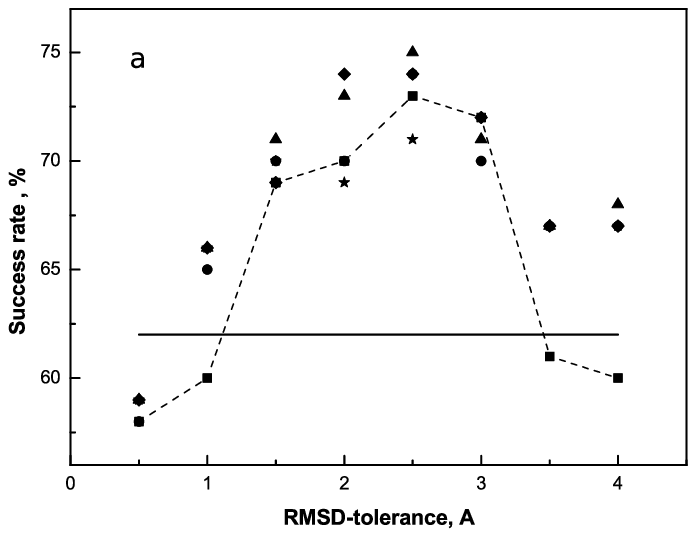}
\\
\\
$\phantom{aaaaaaaaaaaaaaaaaaaaaaaaaa}$\textbf{\large Fig. 2 a}


\newpage

\includegraphics[scale=1.7]{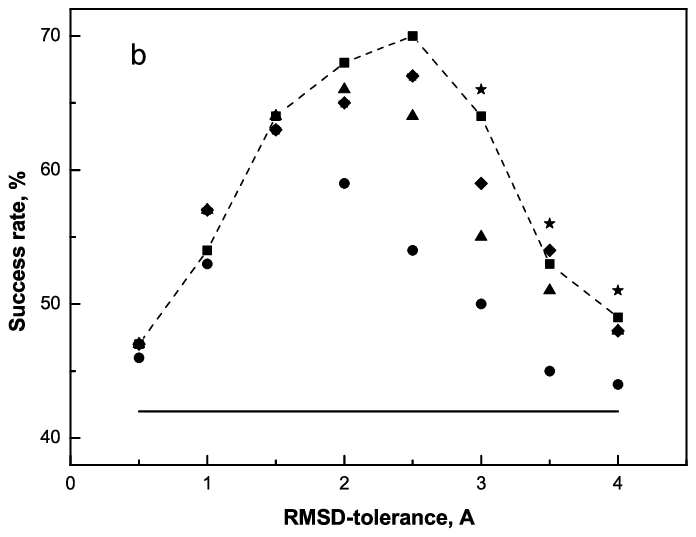}
\\
\\
$\phantom{aaaaaaaaaaaaaaaaaaaaaaaaaa}$\textbf{\large Fig. 2 b}

\newpage
\includegraphics[scale=1.7]{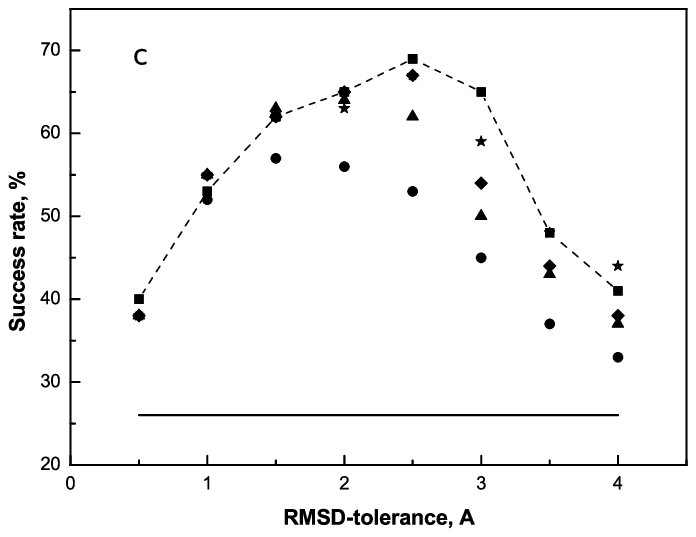}
\\
\\
$\phantom{aaaaaaaaaaaaaaaaaaaaaaaaaa}$\textbf{\large Fig. 2 c}

\newpage
\includegraphics[scale=1.7]{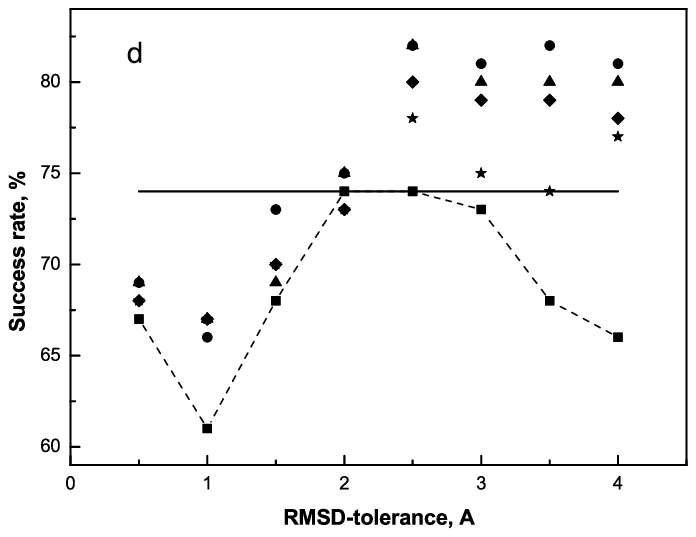}
\\
\\
$\phantom{aaaaaaaaaaaaaaaaaaaaaaaaaa}$\textbf{\large Fig. 2 d}

\newpage

\includegraphics[scale=1.7]{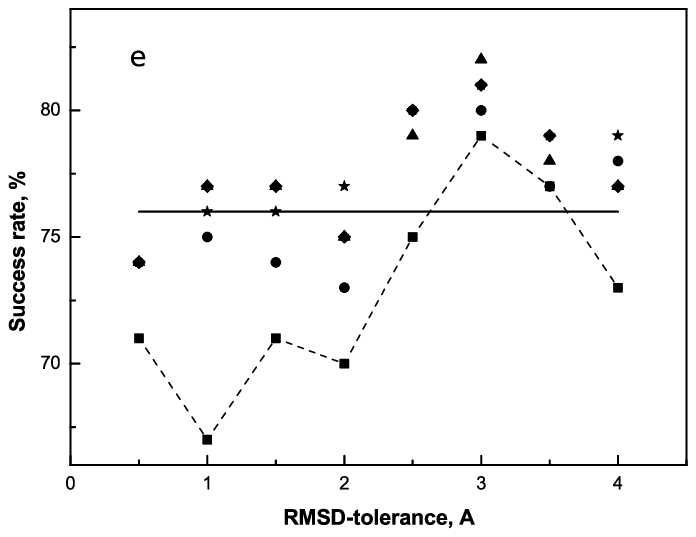}
\\
\\
$\phantom{aaaaaaaaaaaaaaaaaaaaaaaaaa}$\textbf{\large Fig. 2 e}

\newpage

\includegraphics[scale=1.7]{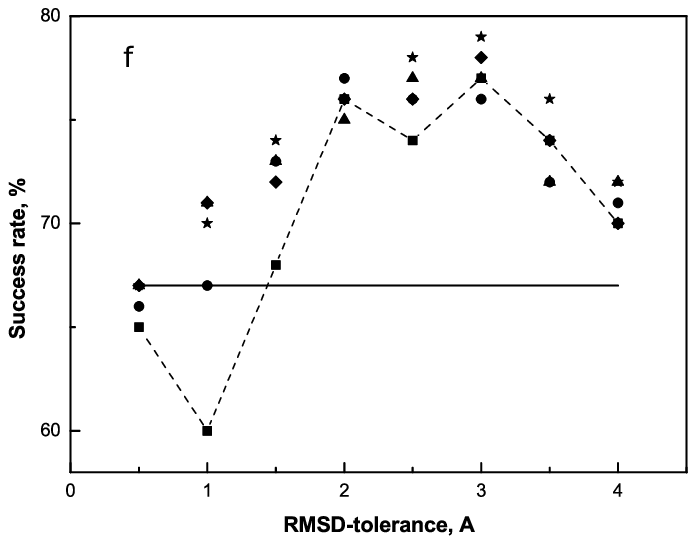}
\\
\\
$\phantom{aaaaaaaaaaaaaaaaaaaaaaaaaa}$\textbf{\large Fig. 2 f}

\newpage

\includegraphics[scale=1.7]{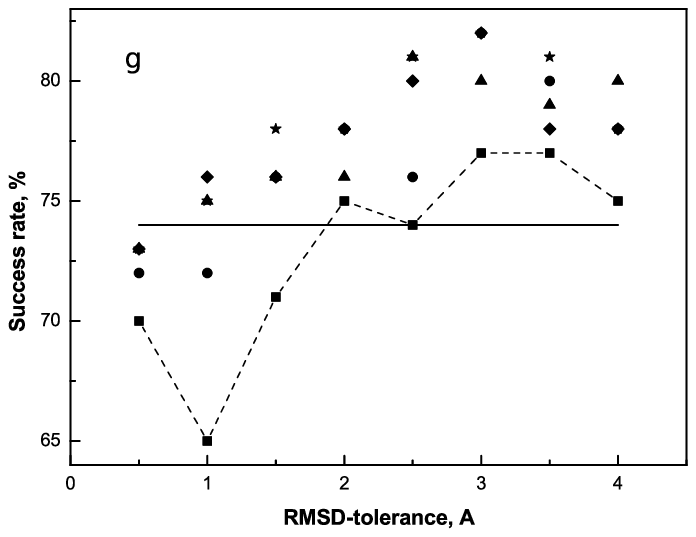}
\\
\\
$\phantom{aaaaaaaaaaaaaaaaaaaaaaaaaa}$\textbf{\large Fig. 2 g}

\newpage

\includegraphics[scale=1.7]{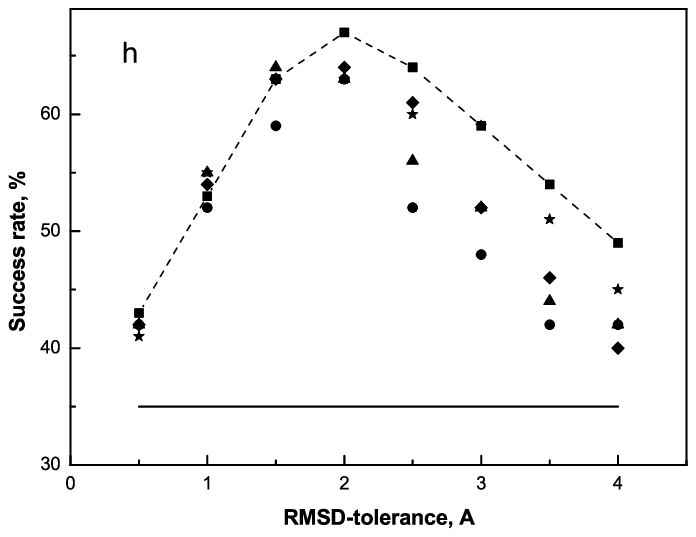}
\\
\\
$\phantom{aaaaaaaaaaaaaaaaaaaaaaaaaa}$\textbf{\large Fig. 2 h}

\newpage

\includegraphics[scale=1.7]{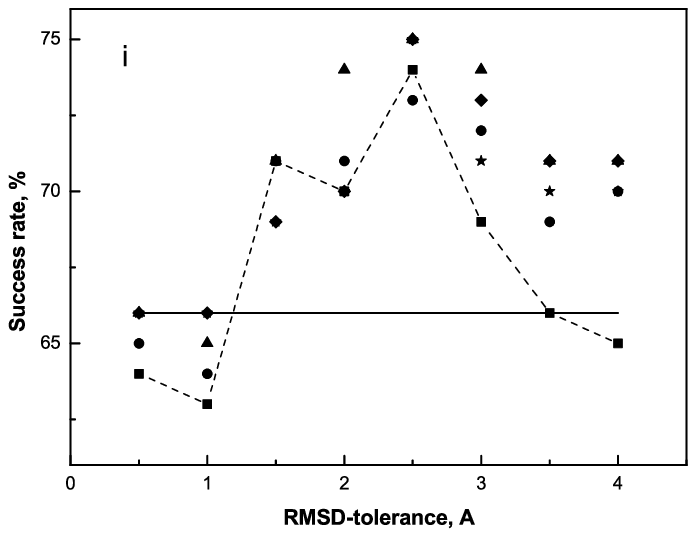}
\\
\\
$\phantom{aaaaaaaaaaaaaaaaaaaaaaaaaa}$\textbf{\large Fig. 2 i}

\newpage

\includegraphics[scale=1.7]{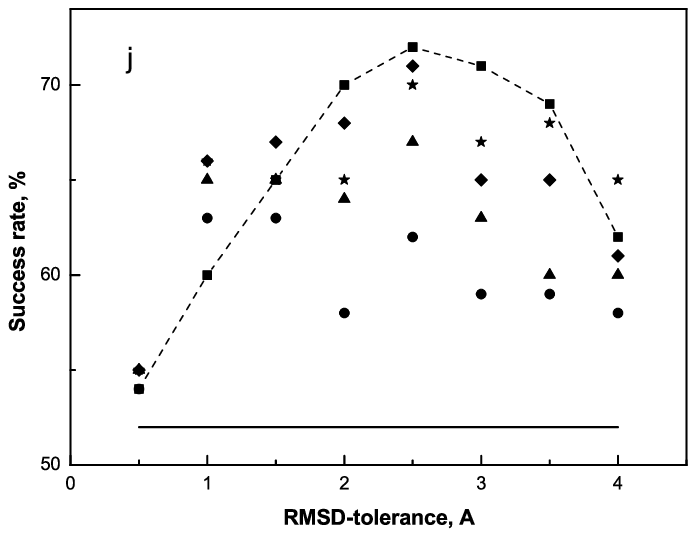}
\\
\\
$\phantom{aaaaaaaaaaaaaaaaaaaaaaaaaa}$\textbf{\large Fig. 2 j}

\newpage

\includegraphics[scale=1.7]{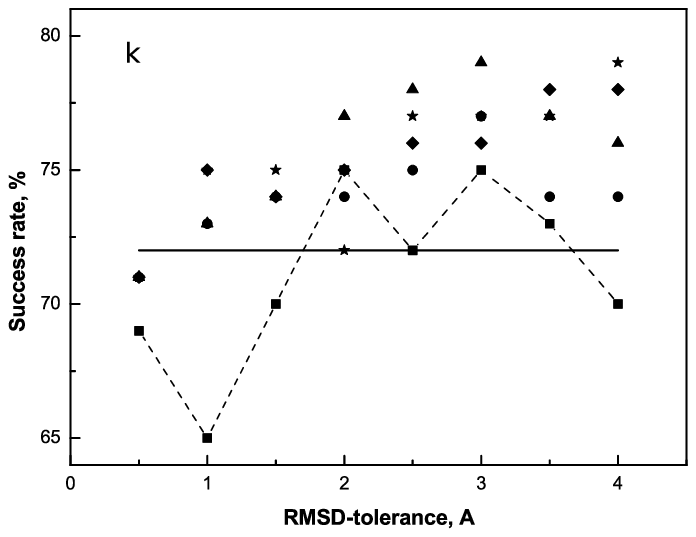}
\\
\\
$\phantom{aaaaaaaaaaaaaaaaaaaaaaaaaa}$\textbf{\large Fig. 2 k}

\end{document}